\documentclass[secnumarabic, pre, showkeys, showpacs]{revtex4}

\usepackage{bm}
\usepackage{amsmath,amsfonts}
\usepackage{color}
\usepackage{graphicx}
\begin{document}

\title{What and how the Michelson interferometer measures}
\author{V.V.Demjanov}
\affiliation{Ushakov State Maritime Academy, Novorossiysk, Russia}
\email{demjanov@nsma.ru}
\date{\today}

\begin{abstract}
Proposed by Maxwell in 1979  detector of aether seems, at a superficial glance, a simple device. For example, Michelson in 1881 thought that he built an instrument that (when you turn it in the horizontal plane) will measure in vacuum (refractive index $n = 1$) the harmonic shift of the interference  fringe, and the presence of air supposedly disturbs measurements. In reality the case is much more involved. Not at once it was understood (the misunderstanding lasted $\sim 90$ years) that the shift of interference fringe occurs only when the carriers of light contain particles, i.e. have $n > 1$. Until now many believe that the ideal environment for the operation of the device is a vacuum, and the shift of the interference fringe observed by some experimenters is produced by systematic errors and the noise because of the chaotic motion of air particles. A reduction of the noise due to pumping the gas from the zones where the light propagates, noticed by some researchers, only enhanced this belief. In 1968-1975 I have demonstrated experimentally that during the evacuation of light paths, i.e. with decreasing the number of particles of the light's carrier, along with the reduction of noise disturbances always necessarily vanishes the harmonic shift of the interference fringe. The latter was found to be detected only in light-carrying media comprising of the mixture of aether with particles of matter. I have mastered by the fringe contrast to single out from much noise and many obstacles the principal interference pattern and was able in a new way to interpret properly the relationship of its contrast with the presence or absence of the fringe shift. After this it became clear when and why the Michelson interferometer confidently detects the absolute velocity of the Earth. It was found to be several hundred km/s.

In the sixth version of this work the slip in formula (\ref{time parallel}) corrected. I consider in more details the major sources of methodical and interpretational failures in experimental detecting the "aether wind" by the Michelson-type interferometers in order to prevent experimenters of their recurrence. I show how to diminish below a measurable level of the sought-for signal the harmful impact of noise and spurious interference that arises because of: the scattering of light by the luminous spot on the semitransparent layer of the bifurcation plate; reflection of light from optical interfaces of composite media; the blurring of the interference fringe because of dissipation in optical media. The reliably reproducible results of measuring the shift of interference shift are presented that was performed in media with refractive index lying in the range $1<n<1.8$.

The questions of interpreting measurements on Michelson interferometers from the very beginning played a decisive role in estimations of results of measurement obtained. In the intentionally incorporated into the fifth version section 7 I analyze the reasons why there developed the belief of  "negativity in principle" of Michelson experiments. Among them are: 1) the lack of understanding of that at $n=1$ interferometer measures nothing except the noises (no fringe shift); 2) exaggerated in $1/\varepsilon\sim 1660$  times expectations of the fringe shift value in the air; 3) understated in $\varepsilon^{- 1/2}\sim 40$ times aether wind speed as evaluated from measured non-zero fringe shifts in the air; 4) undervalued in 19th century more than  $\sim 10$ times expectations of aether wind speed ($30$ km/s instead of $\sim 400$ km/s, that was elucidated only in the 20th century); 5) the lack of knowledge in the 19th century of those new (non-Galilean) principles and measures of relativity, which should be taken into account from the onset in the interpretation of Michelson experiments, but were discovered and began to develop only after 1904.

In the publication Phys.Lett.A 374 (2010) 1110 I informed scientific community about measuring the horizontal projection of absolute velocity of the Earth at the latitude of Obninsk as $140-480$ km/s  depending on the time of day and night. This experimental result became possible only owing to that I was able to overcome the above mentioned methodical and interpretational artifacts.

\end{abstract}
\pacs{42.25.Bs, 42.25.Hz, 42.79.Fm, 42.87.Bg, 78.20.-e}
%phase-shifting interferometry 42.87.Bg
%Interference, optical, 42.25.Hz
%Reflectors, beam splitters, and deflectors 42.79.Fm
%42.25.Bs Wave propagation, transmission and absorbtion
%78.20.-e Optical properties of bulk materials
\keywords{Michelson experiment, optical media, aether wind}
\maketitle

\section{The period 1880 $ - $ 1960}

\begin{figure}[h]
  \begin{center}
\includegraphics[scale=0.6]{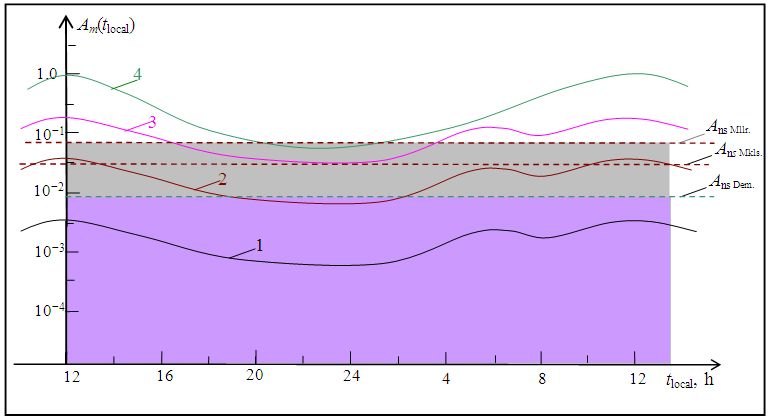}
  \caption{
 The time course (during the day and night) of the relative amplitude $A_m(t_{\textrm{local}})$ of harmonic component $A(\varphi)$ of the shift of interference fringe on the screen of the rotary cross-like interferometer (the local time $t_{\textrm{local}}$), corresponding to 3rd decade of June: 1 $-$ 1881 year, USA, $\sim 42^o$ N, Michelson \cite{Michelson}, $L_\|=L_\perp=1.2$ m (light's carrier is the air); 2 $-$ 1887 year, USA, $\sim 42^o$ N, Michelson$\&$Morley \cite{Michelson Morley}, $L_\|=L_\perp=11$ m (air); 3 $-$ 1926 year, $L_\|=L_\perp=32$ m, USA, Miller \cite{Miller}, $\sim 42^o$ N (air); 4 $-$ 1971 year, USSR, $\sim 55,8^o$ N Demjanov \cite{Demjanov rus}, $L_\|=L_\perp=0.2$ m (fused quartz). $A_{nsi}$ are estimations of noise level of the experimental installations of Miller, Michelson and Demjanov.
}\label{fig1}
\end{center}
\end{figure}

The combined interpretation of all key experiments on the Michelson-type interferometer is presented  in Fig.\ref {fig1} by four time dependencies $A_m (t_\textrm{local})$ of amplitude $ A_m $ of the relative shift of interference fringe taken during the full 24-hour cycle of day and night. In this figure the amplitude $ A_m = X_m / X_o $ of the relative shift of interference fringe is expressed through the real amplitude of the observed fringe shift  in the dimensions  of width $ X_o $ of the fringe. In my experimental setup the interference pattern was visualized by using a home-made camera (with a microscope objective before the vidicon screen) and 18 cm stationary kinescope, where the interference fringe had a width of $ X_o = 90 $ mm.

I managed to reproduce in Fig.\ref{fig1} models of dependencies (1-3), as occurred in the world-famous experiments of Michelson (1881), Michelson$\&$Morley (1887), Miller (1926), which are interpreted as evidence of ''negative'' Michelson-type experiments. They have been transcribed by me in the form of curves (1-3) after my experiments (1968-1974 period), which only recently  became possible to be published \cite{Demjanov rus, Demjanov}. The curve 4 in Fig.\ref{fig1} obtained by me gives the answer, why there was ordained for each of these experiments the fate of being  "negative". Only when there was obtained the amplitude of shift ($X_m>>X_{ns}$) of the fringe many times exceeding the level ($X_{ns}$) of the device noise, it became clear both the unconditional positiveness of the idea by Maxwell how to observe the reaction of aether, and origins of the failure of the first experiments of Michelson in which the measured parameter $ X_m $ turned out to be embraced by all-concealing and suppressing noise $A_{ns}>>A_m$ (see Fig.\ref{fig1}). This became possible only after I was able to understand and discriminate the response of inertial and non-inertial objects in the interferometer, which Michelson in 1881 did not distinguish.

Indeed, according to proposed in 1881 theoretical model of processing experimental results, based on the classical rule of composition of velocities ($ c\pm V $) of the  wave of light ($ c $), as non-inertial object, and of the inertial light source and light's carrier moving steadily with velocity $ V $ in stationary aether, Michelson obtained the following formula to determine the speed of "aether wind", neglecting the phenomenon of drift of the beam in the arm $L_\perp$  (this inaccuracy was corrected afterwards by Lorentz, see at length in section 7):
\begin{equation}
V=c [A_m\lambda/(2L)]^{1/2}\label{aether wind}.
\end{equation}
In anticipation of the experiment, Michelson estimated  from formula (\ref{aether wind}) the expected value of the amplitude of the relative shift by its reversal relative to $A_m$:
\begin{equation}
A_{m\,\,\textrm{exp.}}=2LB^2/\lambda\label{shift}.
\end{equation}
In (\ref{aether wind}) and (\ref{shift}) there are designated: $c$ is the speed of light in vacuum; $V$ speed of the experimental setup (the light source and particles of the light's carrier) relative to aether; $B=V/c$; $\lambda$ is the wavelength in vacuum, and $A_m=X_m/X_o=c\Delta t/\lambda$. Note that neither the well-known in those years refractive index $n$ of optical medium, whose role in Michelson experiment played the air ($ n = 1.0003 $), nor less known then permittivity $\varepsilon=n^2=1.0006$ of the air were not taken into account in formula (\ref{aether wind}).

Expectations of Michelson in 1881 were optimistic. For $L_\|=L_\perp=1.2$ m and $B^2=10^{-8}$ (i.e. for the linear velocity of the Earth in its orbit around the Sun $\sim 30$ km/s) at the beam of visible light he expected to obtain the amplitude $A_{m\,\,\textrm{exp.}}=0.04$ [3] corresponding to the shift of the fringe by 1/25 of the bandwidth. In evaluating the resolving power of his interferometer (1/40 of the bandwidth \cite{Michelson}) he was sure that he detects a shift of the fringe (because the ratio signal/noise was expected to be $\sim 2$).  When he made measurements, he found no indications of the shift of interference fringe (i.e. he obtained $A_m=0$). Therefrom the world-famous ''negative'' conclusion was drawn that there is no aether.

In fact, this was only the starting point of an intricate scientific problem lasting  for the period of a century: what and how does the Michelson interferometer measure? The real ''picture of non-observability'' of the amplitude $A_m$ of relative shift  of the interference fringe in Michelson's  1881 experiment, from the height of my current understanding of the problem, looked as shown in Fig.\ref{fig1} curve 1, i.e. the sought-for in the experiment value of $A_m$ was sunk in the noise, the intensity of which exceeded it hundreds of times.

Perhaps realizing this, in 1887 Michelson$\&$Morley \cite{Michelson Morley} increased the length of the interferometer arms to from $L_\|=L_\perp=1.2$ to 11 meters (almost 10 times compared with the experiment of 1881 year). By formula (\ref{shift}) with $L_\|=L_\perp=11$ m and $B^2=10^{-8}$ they expected to obtain $A_{m\,\,\textrm{exp.}}=0.4$ \cite{Michelson Morley}. The shift almost 1/2 of the bandwidth is impossible to be unnoticed. However, they again found no shift of interference fringe. Actually, this ''picture'' of non-observability of relative amplitude of the shift $A_m$ of interference fringe was lost in the noise, with only difference that now the noise exceeded the required shift not in hundreds, but in dozens of times (curve 2 in Fig.\ref{fig1}). But neither Michelson$\&$Morley nor  other scientists until 1926 year had no idea about this. Experiments of 1881 and 1887, indeed, should be considered as ''negative'' if  interpreted with formulas (\ref{aether wind}) and (\ref{shift}).

Only having obtained thoroughly cleansed from the noise curve 4 in Fig.\ref{fig1} I realized that in the experiment of Michelson$\&$Morley only 1 hour per day there evolved signal/noise ratio, which is close to unity ($A_m/A_{ns}\sim 1$), and then in the other 23 hours again it decreased tenfold (see the run of the curve in Fig.\ref{fig1}).  Despite of these unfavorable conditions, when no one knew the reasons for discrepancies of expectations $A_{m\,\,\textrm{exp.}}$ due to model (\ref{shift}) and measurements  $A_{m\,\,\textrm{measur.}}$ of the fringe shift in the experiment, Michelson$\&$Morley announced that they were able to briefly record the fringe shift by $\sim 1/30$ of the bandwidth against the background of a big noise \cite{Michelson Morley}. Naturally, this time no one believed it because of the unclear reasons of a  multiordinal discrepancy between the expected value of $A_{m\,\,\textrm{exp.}}\sim 0.4$ according to the model (\ref{shift})  and measured value: $A_{m\,\,\textrm{measur.}}\sim 0.03$. This lasted until 1926.

In 1926  Miller repeated the experiment of Michelson using interferometer with greater length of arms, $L_\|=L_\perp\sim 32$ m. The picture, which he first observed during almost the whole day and part of the night, presented in Fig.\ref{fig1} curve 3. I calculated it by formula (\ref{shift})  from the data of Miller \cite{Miller} on the velocity of ''aether wind'' that he obtained  processing his measurements by formula (\ref{aether wind}). Here we see that the experiment is already comfortable with extracting out of the noise the amplitude of the fringe shift (with the ratio $A_m/A_{ns}\sim 2-3$) in a half-day and night. Yet, since there remained unknown the reason why in the experiment of Miller the anticipated  by formula (\ref{shift}) fringe shift promised to give $A_{m\,\,\textrm{exp.}}=1.2$, and in fact at ''peak'' it gave $A_{m\,\,\textrm{measur.}}\sim 0.1$, his results were also listed as ''negative''. Such conclusion was instigated by  new knowledge disclosed in twentieth century.

Astronomical observations of those years have demonstrated that the velocity of the Earth in space is determined not only by the linear velocity ($\sim 30$ km/s) of its rotation in its orbit around the sun, but by the order of magnitude greater linear velocity ($\sim 300$ km/s) of its rotation in its orbit around the Galactic center \cite{Shklovsky}. This corresponds to the parameter $B^2=V^2/c^2\sim 10^{-6}$. If Michelson had recognized this in 1881, he would obtain by (\ref{shift}) the estimation of $A_{m\,\,\textrm{measur.}}$ not 0.04, but 4.0 (i.e. the fringe shift of four bandwidths!). Such a reaction can not be overlooked. To an even greater extent there would alarmed all the discrepancy between the expected fringe shift ($A_{m\,\,\textrm{exp.}}\sim 40$ for $B^2\sim 10^{-6}$) and  obtained by Michelson$\&$Morley in 1887 experiment  value $A_{m\,\,\textrm{measur.}}\sim 0.1$ (for $L_\|=L_\perp=11$ m). Overstatement of  $A_{m\,\,\textrm{exp.}}$ in thousand times, that I was capable to communicate only in \cite{Demjanov rus, Demjanov}, may have prompted theorists to think over the problem of discrepancies in mathematical processing of the experiment.

In the end, the final sentence of the Michelson experiment as ''negative'' was contributed by three circumstances. \textit{First}, experiments were not reproducible and were not confirmed by laboratory measurements in  vacuum \cite{Joos}, i.e. for $ n = 1 $ (even though everyone thought that this is the most sterile conditions to detect the ''aether wind''). Now, after reporting my experiments in \cite{Demjanov rus, Demjanov}, it became clear that it was erroneous estimations based on a misunderstanding of the principle of the interferometer. \textit{Second}, the experiments were not reproducible and not confirmed because of the ignorance of how the shift of the fringe (detected in the rotation of the optical platform of the device in the horizontal plane) changes depending on the time, the day or night,  of shooting (see Fig.\ref{fig2}). Only after that to the 1971 year I had found such a relationship by analyzing the patterns of seasonal drift of the curve $A_m(t_{\textrm{local}})$ during the year (see Fig.\ref{fig2}), all measurements at any date and any time of day or night became stably reproducible. \textit{Third}, even taking into account all the above-mentioned sporadic cases of detection by experimenters of the relative amplitudes of  shift $A_{m\,\,\textrm{measur.}}$, their treatment by the Michelson  formula (\ref{aether wind}) invariably gave the ''aether wind'' speed $5<V<10$ km/s \cite{Michelson Morley, Miller, Shamir}. And this is tantamount to its absence, especially after it came to notice that the Earth is rushing in space relative to the stars in our galaxy at a speed not less than 300 km/s \cite{Shklovsky}.

Such were dramatic attempts to measure the speed of ''aether  wind'' with Michelson interferometer to the beginning of 1960. It is indicative that they do not distinguish between measurements on interferometers with air or vacuum atmosphere in the zones of rays, but it was believed that the results from the evacuated zone of propagation of rays should be trusted more. This characterizes the stage of history of science considered as a period of complete lack of understanding the physical principle of Michelson interferometer.

\section{The period of 1960 $ - $ 2010}

\begin{figure}[h]
  \begin{center}
\includegraphics[scale=0.6]{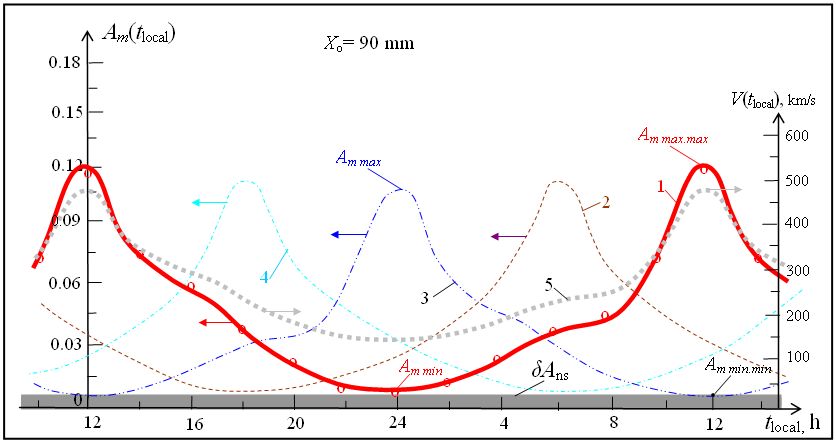}
  \caption{Patterns (1-4) of the seasonal displacement of the round-the-clock dependencies $A_m(t_\textrm{local})$ of the observed relative amplitude $A_m$ of the harmonic component $\Delta A(\varphi)$ of the interference fringe shifts on the kinescope's screen when turning the platform of the rotary cross-like interferometer with water light's carriers only in the horizontal plane, which were measured every two hours during the day and night Moscow time in the third decade of the month: 1 $-$ June; 2 $-$ September; 3 $-$ December; 4 $-$ March (based on measurements of  1969$-$1971 years); 5 $-$ the velocity of ''aether wind'', computed from the curve 1 by formula (3). The measurements were performed on water light's carriers ($\textrm{H}_2\textrm{O}$ distillate in glass tubes) with following parameters of the interferometer: $L_\|=L_\perp=0.17$ m, $\lambda=9\cdot10^{-7}$ m, $\Delta\varepsilon_{\textrm{H}_2\textrm{O}}\approx0.83$; $X_o=90$ mm is the width of the fringe on the screen of the kinescope; $\delta_{ns}$ the level of jitter noise of the interference pattern. $A_{m\,\,max}$ and $A_{m\,\,min}$ is the maximal and minimal amplitude of the fringe shift for 24-hour period of measurement in day and night; $A_{m\,\,max\,\,max}$ is the largest (in the year) maximum amplitude of fringe shifts observed from 18 to 25 June each year; $A_{m\,\,min\,\,min}$ is the smallest (in the year) minimum amplitude of fringe shifts observed from 18 to 25 December each year (all for the latitude of Obninsk).}\label{fig2}
\end{center}
\end{figure}

In 1960th years there appeared first attempts of measuring on Michelson interferometer with high optical density media as light's carriers \cite{Demjanov rus, Demjanov, Shamir}, and in subsequent decades, until now, interest in them is ever growing \cite{Shamir, Shamir Lipson, Trimmer, Cahill Kitto, Cahill}. My measurements showed (see curve 4 in Fig.\ref{fig1} and curve 1 in Fig.\ref{fig2}) the enormous potentiality to improve the signal (in the interferometer such a signal is $A_m$) raising it over its own noise ($\delta A_{ns}$). Their realization enabled me to respond constructively to the above three questions. Hourly and monthly view of changes in the amplitude of shift of the interference fringe (with turning the interferometer in the horizontal plane) for the signal/noise ratio not less than  $1.5-15$ during the whole year is shown in Fig.\ref{fig2}.

It shows the measured by me four dependencies $A_m(t_{\textrm{local}})$ in four characteristic times of the year, specially referred to the third decade of the month. These data are typical for the latitude of Obninsk ($ 55,80^o $ N). Curves were obtained in: 1 $-$ June, 2 $-$September, 3 $-$December, and 4$-$ March. Measurements were carried out in 1969 and checked several times until 1974 year. From these dependencies I have drawn for the first time the following regularities in the drift of the position of the ''peak'' of $A_m(t_\textrm{local})$ along the axis of the local time:

$-$  maximum maximorum of the amplitude of the relative shift of the fringe $A_{m\,\,max\,\,max}$ (when turning the interferometer in the horizontal plane) is observed in the range $11^{30}-12^{30}$ o'clock local time from 18th to 24th June each year, à minimum minimorum $A_{m\,\,min\,\,min}$ is observed in the range $11^{30}-12^{30}$ o'clock in  December each year,

$-$  maximum amplitude of the relative fringe shift $A_{m\,\,max}$ is displaced by two hours in each month in the order of the adopted numeration of months (to the left in Fig.\ref{fig2}); thus, the maxima of curves 1, 2, 3 and 4, taken with the step three months, moved relative one to another next on the adjacent curve by 6 hours;

$-$  ratio of the amplitude $A_{m\,\,max\,\,max}$ of the fringe shift, measured at 12 o'clock June 22, to the amplitude  $A_{m\,\,max}$ of the fringe shift, measured at 24 o'clock December 22, was equal to $\sim 1.06$ (at the latitude of Obninsk);

$-$ calculated from the dependence $A_m(t_{\textrm{local}})$ by formula (\ref{aether wind Demjanov}) the value of the component of velocity ($V$) of ''aether wind'', as shown by curve 5 in Fig.\ref{fig2}, changes at the latitude of Obninsk in the range of values  $140<V<480$ km/s.

Thus, experimental results obtained by me indicate that there exist reliable methods to detect the fringe shift on  Michelson-type interferometer with the signal/noise ratio more than 10. With such a certainty of execution of   experiments on Michelson-type  interferometers and those detailed measures of modernization with the help of high optical density  media as light's carriers, which I have proposed and experimentally verified \cite{Demjanov rus}, it would seem no reason to believe that Michelson-type experiments are ''negative''. Clearly, in order to confirm it, my measurements and their new interpretation should be revised and reproduced by other researchers.

However, here there may again repeat the story of the previous period. Then a misunderstanding of the true physical principle of Michelson interferometer carried off experimenters on a wrong way of remaking Michelson$\&$Morley and Miller experiments, held in the air, supposedly in a more ''pure'' vacuum atmosphere condition of propagating the  rays. Naturally, the results obtained in vacuum gave ''zero'' fringe shifts. This contributed to a denial of the positive results of measurements in air, obtained by Michelson$\&$Morley and Miller. I experimentally disclosed the intrigue of the century-old scientific mistakes and built up a theory to explain them \cite{Demjanov rus, Demjanov}.

Now that I have described \cite{Demjanov rus, Demjanov} the positive results concerning the sharp increase in sensitivity to ''aether wind'' of the interferometer with high optical density light's carriers, as I foresee, my results may not be supported by hasty attempts to repeat them in facilities that do not exclude methodical artifacts of spurious interference, noise-polluting fringe shifts. In particular, the monograph  \cite{Cahill}  reported that in \cite{Shamir} there were obtained almost zero shifts (1/3000) of the fringe on the interferometer with high optical density light's carriers. The authors \cite{Shamir} gave to such a shift of the fringe a controversial interpretation according to which the speed of aether wind is $\sim 6$ km/s (i.e. against the background of 300 km/s it is almost zero). Surely, orthodoxies of the official science and its ''principle of relativity'' believing the ''negativeness''   of Michelson experiments to be its experimental foundation, enthusiastically greet reports of experiments which failed to detect the expected shift of interference fringe corresponding to an estimation of the aether wind velocity a few hundreds km/s. While they meet with hostility the reports on any positive results of the experimental registering the finite shift of interference fringe and drawing from it the right order of the aether wind velocity (several hundreds km/s).

Since after 1969 no new information on measurements of the authors \cite{Shamir, Shamir Lipson} appeared, and published by me \cite{Demjanov rus, Demjanov} the positive results of those same years will take a time for them to be verified, I will impart to experimenters my experience how to overcome the difficulties of measuring on Michelson-type interferometer with different optical media, including  laboratory vacuum. I will describe few methodical artifacts that may lead to a false apparent absence (or drastic underestimation of the value) of the sought-for shift of interference fringe in the devices, which require the use of optical-transparent containers (for liquids and gases) or optical-transparent rods. Application of the described below technique of removing artifacts is given in works \cite{Demjanov why Shamir, Demjanov why Trimmer} where I explained properly the mistakes made by authors \cite{Shamir, Trimmer}, fixed them, and from measured by these authors fringe shifts ($A_m=1/3000$ \cite{Shamir} and $A_m=10^{-5}$ \cite{Trimmer}) obtained a consistent estimation of the aether wind velocity $V\sim 400$ km/s.

\section{Positiveness of Michelson-type experiments}

\begin{figure}[h]
  \begin{center}
\includegraphics[scale=0.6]{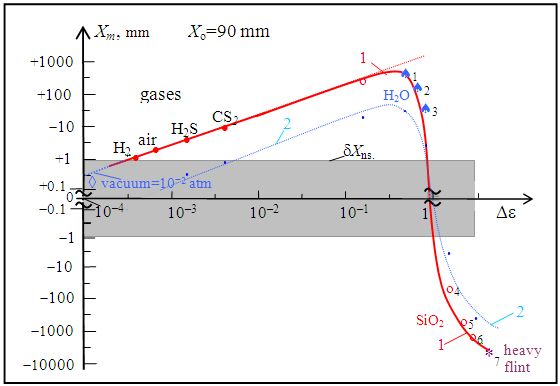}
  \caption{Dependence $X_m(\Delta\varepsilon)$ of the amplitude $X_m$ of interference fringe shift, as observed at the screen of the kinescope, on the contribution $\Delta\varepsilon$ of particles into the full dielectric permittivity  ($\varepsilon=1+\Delta\varepsilon$) for various light's carriers: ${\color{blue}\diamondsuit}$  $-$ vacuum, $10^{-1}$ atm; ${\color{red}\bullet}$ gases; ${\color{blue}\spadesuit}_i$ water; ${\color{red}o}_j$ fused quartz; ${\color{magenta}*}_7$ heavy flint glass at a blue ray (all experimental values are reduced to $L_\|=L_\perp=6.0$ m and $\lambda =6\cdot10^{-7}$ m). Curve 1 corresponds to $X_{m\,\,max}$, and curve 2 to $X_{m\,\,min}$ in the notations of Fig.2; at the local time of observation $X_{m\,\,max}$ is correspondent by the projections (about $480$ km/s) of the ''aether wind'' velocity on the horizontal plane of the device, and for $X_{m\,\,min}$  these projections  decrease to about $140$ km/s (at the latitude of Obninsk).  $X_o=90$ mm is the width of the interference fringe, $\delta X_\textrm{ns}$ jitter noise of the interference fringe on the screen of the kinescope.}\label{fig3}
\end{center}
\end{figure}

Fig.\ref{fig3} shows my results of measurements on the Michelson interferometer published in \cite{Demjanov rus, Demjanov}. Here refractive index is expressed in terms of the optical dielectric permittivity $\varepsilon=n^2$ as   characteristic of Maxwell's electrodynamics of continuous media. In Maxwell's theory the value ($n^2-1$) got (back to  1870) a clear physical sense of $(\varepsilon-1)=\Delta\varepsilon$, under which the total permittivity of optical-transparent  medium is always comprised of the sum of the polarization contribution of aether (\textbf{1.}) without particles and contribution of the polarization of particles $\Delta\varepsilon$, always present in aether, i.e. $\varepsilon=\textbf{1.}+\Delta\varepsilon$. I found that the amplitude of interference fringe depends not on the entire  dielectric permittivity of light's carriers of the interferometer, but only on its part $\Delta\varepsilon$ which is due to the polarization contribution of inertial particles of light's carriers, and completely independent on the polarization contribution of non-inertial aether that is equal to \textbf{1.} Along with this I have found that the thermal noise of polarization of optical media grows proportionally to its dielectric permittivity ($\varepsilon=1+\Delta\varepsilon$). So, I realized that sensitivity of Michelson interferometer with air light's carriers ($\Delta\varepsilon=0.0006$) can be enhanced approximately by three orders of magnitude using liquid and solid optical media with $\Delta\varepsilon\sim 3$. Indeed the fringe shift in the solid light's carriers rises $3/0.0006$ times comparing with air, and thermal noises of these media increase only $\varepsilon=1+3=4$ times.

In Fig.\ref{fig3} the dependence of the fringe shift amplitude $A_m(\Delta\varepsilon)$ is represented on $\Delta\varepsilon$ axis in logarithmic scale. This was done deliberately in order to show the new experimental points obtained not only in the air but also in other gases including laboratory vacuum. It was impossible to show them in the figure of \cite{Demjanov}, where I applied the linear scale of $\Delta\varepsilon$. The pattern in Fig.\ref{fig3} covers almost the whole area  ($1<\varepsilon<3$) of known to science optical-transparent media (from a laboratory vacuum and different gases to liquids and solids) which were tried by me as light's carriers in Michelson interferometer. But the main point is not even in that.

As will be shown below, passing from measurements of interference fringe shift on Michelson interferometer with an air atmosphere to the interferometer, where in the zone of its orthogonal rays the air is replaced by other media, there appear many methodical risks of arising systematic errors that could give false appearance of the absence of shift of interference fringe. All of them are surmountable. I have no doubt that having overcome all methodical difficulties and taking into account my recommendations, or using new methods, that perhaps I do not know, each experimenter will be able to reproduce the dependence  $A_m(\Delta\varepsilon)$ disclosed by me in Fig.\ref{fig3}.

All the results presented in Fig.\ref{fig3} are reduced intentionally to the data obtained on an interferometer with air light's carrier ($\Delta\varepsilon_{air}=0.0006$). This is done in order to show the tremendous growth of the sensitivity and resolving power of the interferometer equipped by  light's carriers with much greater magnitude of $\Delta\varepsilon$  than that of the air. In addition, in Fig.\ref{fig3} these results are more comprehensively represented in the range of permittivity of optical media $1<\varepsilon <1.01$. On the one hand, they clearly show that  Michelson interferometer in the absence of material light's carriers (i.e. in vacuum at  $\varepsilon=\textbf{1.}$, when $\Delta\varepsilon=0$) is not sensitive to the kinetic interaction with aether, i.e. gives $A_m=0$. On the other, they completely refute  \cite{Demjanov rus, Demjanov} the myth of the ''negativeness''  of Michelson experiments $-$ already at $\Delta\varepsilon>0.001$, i.e. at concentration of particles in light's carrier $>10^{19}\,\,\textrm{cm}^{-3}$.

Finding the linear dependence of $X_m$ on $\Delta\varepsilon$ in the range $0.000006<\Delta\varepsilon<0.01$ enabled me by the back extrapolation to draw a conclusion above mentioned concerning the loss of interferometer's sensitivity to the shift of interference fringe when $\varepsilon\rightarrow  1$. Thus, in an experiment by Fig. \ref{fig3} with $L_\|=L_\perp=6$ m in the air ($\Delta\varepsilon=0.0006$) in the ''rush hour'' of the maximum of the observed amplitude of the fringe shift $A_m$ (by Fig.\ref{fig2} this is $t_{\textrm{local}}=12$ o'clock) there is registered the fringe shift $X_m\sim 2$ mm. When I started to pump out the air from the zones of propagation of rays of the interferometer, already at the residual pressure $\sim 0.1$ atm in flasks (i.e. when $\Delta\varepsilon\sim 0.00006$) the fringe shift ceased to be observed (it seems  that $X_m\sim 0$), and this is with the resolving power of my apparatus $\sim 0.7$ mm. Fig.\ref{fig3} shows by the point $\diamondsuit$ that in reality the fringe shift   exists at $\Delta\varepsilon\sim 0.00006$, but it is buried in the noise of the device, the level of which was $\delta X_{ns}\sim 0.7$ mm. Let us reverse the direction in which  $\Delta\varepsilon$ to be varied. Raising the air pressure in flasks up to 3 atm the value of $\Delta\varepsilon$ will rise to about 0.0018, while the shift of the fringe will increase to  6 mm.  Replacing the air in flasks by carbon disulfide whose optical permittivity is $\varepsilon_{\textrm{CS}_2}=1.0036$, we get almost a sixfold increase in the amplitude of the fringe shift: $X_{m\,\textrm{CS}_2}\sim 11$ mm (Fig.\ref{fig3}). These data formed the basis for  empirical refinement of Michelson formula (\ref{aether wind}) in the following original form \cite{Demjanov rus}: $V\approx c [A_m\lambda/(L\Delta\varepsilon)]^{1/2}$.

Performing experiments on the interferometer with light's carriers in the range $1.1<\varepsilon<3.0$ I have found a violation of linearity at $\Delta\varepsilon>0.1$. Further, it became clear that in the whole the curve $A_m(\Delta\varepsilon)$ is S-shaped and changes its sign at $\Delta\varepsilon=\varepsilon-1=n^2-1=1$. To interpret this dependence I have proposed (in 1971) the model for determining the rate of "ether wind" (the details of its derivation are described in \cite{Demjanov rus, Demjanov}):
\begin{equation}
V=c\left[\frac{nA_m\lambda}{2L\Delta\varepsilon(1-\Delta\varepsilon )}\right]^{1/2}\label{aether wind Demjanov}
\end{equation}
Reversing (\ref{aether wind Demjanov}), a new formula for assessing the expected value of the relative amplitudes $A_m$  of the harmonic shift of the fringe was obtained:
\begin{equation}
A_{m\,\,\textrm{exp.}}= \frac{2B^2L\Delta\varepsilon(1-\Delta\varepsilon)}{\lambda n}\label{shift Demjanov}
\end{equation}
that for gases ($\Delta\varepsilon\ll1$) really gives the linear dependence $A_m(\Delta\varepsilon)$ which at $\Delta\varepsilon\rightarrow 0$ tends the fringe shift to zero $[A_m(\Delta\varepsilon\rightarrow 0)\rightarrow 0]$. The former formula (\ref{shift}) described none of this.

These formulas explain all known since 1881 year results of experiments on  Michelson interferometer and all vicissitudes of its misinterpretation:

1) The ''lack'' of fringe shift in experiments at normal air pressure with interferometer arms $L_\|=L_\bot=L<5$ m \cite{Michelson}. With the resolving power of the interferometer $\sim 1/40$ of the bandwidth  the expected by  (\ref{shift Demjanov}) fringe shift is obtained much less than  1/40, i.e. there is a shift, but it is not observed being buried in the noise.

2) In the experiments  at  normal air pressure with interferometer arms $L_\|=L_\bot=L>20$ m \cite{Michelson Morley, Miller} a fringe shift is detected ($A_m\neq 0$), but the results were processed with the Michelson formula (\ref{aether wind}), i.e. wrongly. In this case, the speed of ''aether wind'' is invariably  underestimated giving values  in the range $5<V<10$ km/s. I have found that non-accounting in (\ref{aether wind}) the dielectric permittivity of the air light's carrier in Michelson interferometer gives hundred-fold underreporting of the velocity of ''aether wind'' by  (\ref{aether wind}) in comparison with the correct formula  (\ref{aether wind Demjanov}). Clearly, that obtained by Miller with (\ref{aether wind}) estimations of the velocity of ''aether wind'' $5<V<10$ km/s were considered as ''noise'' of the device.  Processing Miller  experiments (curve 3 in Fig.\ref{fig1}) by formula (\ref{aether wind Demjanov}) gives the correct estimation of the speed of ''aether wind'': $200<V<400$ km/s.

3) Finite interference fringe shifts ($A_m\neq 0$) were found in \cite{Michelson Morley}, by Miller  \cite{Miller} and others on air interferometers. All "refutations" of them on facilities where the zones of light propagation were vacuumed (for example, \cite{Joos}) are based on misunderstanding the operating principle of the interferometer. The point is that any laboratory "vacuum" is really not a true vacuum since has $n>1$. I proved, by means of gradual pumping the air from the zones of light propagation,  that with $\varepsilon\rightarrow 1$ ($\Delta\varepsilon\rightarrow 0$) the sensitivity of interferometer to ''aether wind'' is reduced to zero ($A_m\rightarrow 0$). But in vacuuming the beams propagation zones,  $A_m$ always decreases so times as $\Delta\varepsilon$ decreases, i.e. the ratio $A_m/\Delta\varepsilon$ retains constant. Keeping the ratio $A_m/\Delta\varepsilon=$const in experiments with gases and in laboratory vacuum is evident immediately from (\ref{shift Demjanov}): $A_m/\Delta\varepsilon=2B^2L/\lambda=$const, where $B=V/c\sim 1.5\cdot 10^{-3}$ is the constant of the Earth's motion in aether with the velocity $V$ at least $V\sim 400$ km/s, and $L/\lambda$ the constant of the experimental device.

Publishing in  \cite{Demjanov} my results, obtained on Michelson-type interferometers with liquid and solid media  as light's carriers, will likely revive attempts to repeat them, verify, check and recheck. Formulas (\ref{aether wind Demjanov}) and (\ref{shift Demjanov})  provide the necessary basis for a definitive debunking the myth of ''negativeness'' of Michelson experiments. However, the experience gained by me in measuring the amplitude of the shift of interference fringe, when installing into the device light's carriers made of different optical-transparent  media, tells me that there may appear ''negative'' results in the recurrence of my measurements. While I am aware of only three research groups \cite{Shamir, Shamir Lipson, Trimmer, Cahill Kitto, Cahill} stating that measurements on Michelson interferometer with solid light's carriers (plexiglas and fused quartz) gave ''negative'' result (i.e.  not gave the expected large shift of interference fringe, which would be many times higher than the noise of the device). Below I will discuss some hidden reasons that can lead to failure of all measurements on interferometers with liquid and solid carriers of light having rectangular ends of cells or rods and ñ nonoptimal (by the noise) choice of the point where back rays are brought together for interference.

\section{Two secrets of successful measurements on Michelson-type interferometer with liquid and solid optical-transparent media as light's carriers}

I will briefly describe here two probable reasons, the neglect of which may nullify the results of measurements on all the expected high sensitivity (by the shift of interference fringes) Michelson-type interferometers with hight optical density light's carriers made of optical-transparent liquids or solids. The point is that in the traditional scheme of the interferometer light's rays propagate in a homogeneous air without encountering any other optical-transparent media with different refractive index. When you put a high optical density media across the beam path there arise inevitable unwanted reflections of the part of the beam's energy that may completely disrupt the planned process of basal interference of two main orthogonal beams.

\begin{figure}[h]
  \begin{center}
\includegraphics[scale=0.6]{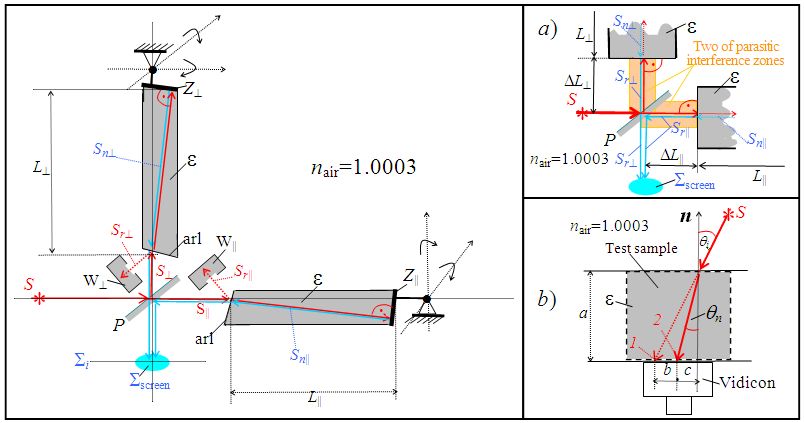}
  \caption{The scheme of rays paths in a Michelson-type interferometer with optical media contained in cells (in case of replacing the air by vacuum, other gases or fluids) or mounted as rods. The construction shown enables us to  eliminate a spurious interference (that obstructs the observation of the basal interference pattern) by means of  beveling the ends of the cells or rods and using W $-$ the trap of reflected waves, Z $-$ the mirror, \textbf{arl} $-$ antireflection layer:
  a) the zone of localization of parasitic interference in the wrong (normal to the rays) orientation of the butt face of a cell or rod;
b) experimental determining the refractive index $n$ (or optical permittivity $\varepsilon=n^2$) by measuring the incidence $\theta_i$ and refraction $\theta_n$ angles via the projection on the vidicon's screen of the light's beam used in interferometer.}\label{fig4}
\end{center}
\end{figure}

Below I discuss (see Fig.\ref{fig4}) how to construct properly the optical scheme of the interferometer so that there will be canceled all possible loci of arising noise-polluting interference, at the same time retaining the basal valid interference of the rays passed along \textit{all the arms} of the interferometer.

Thus, \textbf{the first possible cause} of obtaining a negative result (the "null" shift of interference fringe) is concerned with the ousting the faint interference pattern of basal rays, attenuated by the passage through all the length of the interferometer's light-carrying medium, by the more intense and contrasting spurious interference, arising because of reflections from the interface at the ends of the rods or cells, which are usually kept rectangular,  where the jump of the refractive index takes place.

Fig.\ref{fig4} shows two pairs of these competing zones of the interferometer. The first zone produces the main interference pattern by means of the rays $S_{n\|}$ and $S_{n\perp}$ attenuated by the double path through the optical medium of the arms having considerably greater extinction in comparison with the air.  The second zone produces the stray interference pattern formed by the rays  $S_{r\|}$ and $S_{r\perp}$ reflected from the ends of the rods (cells) at short ($\sim 1$ cm) distances $\Delta L_\|$ and $\Delta L_\perp$ in the air between the plate $P$ and the ends of the optical material. These rays are not attenuated in any way. The dominance of the spurious pattern over the basal one is concerned with the orthogonality of the rays entering the butt of the optical material.

The ratio of the intensity of relevant rays $S_{n\bot}$ and $S_{n\|}$ to the intensity of the spurious ones $S_{r\bot}$ and $S_{r\|}$ (see Fig.\ref{fig4}a), as a rule, even in highly transparent media is  $S_{n\bot}/S_{r\bot}\leq1$ and $S_{n\|}/S_{r\|}\leq1$, and in media with damping $\sim 0.1-1.0$ dB/cm and greater even with  $L_\|=L_\perp\sim 20-30$ cm there is at all $S_{n\bot}/S_{r\bot}<0.1$ and $S_{n\|}/S_{r\|}<0.1$. This is a situation almost similar to that occurred 100 years ago, when the results of measuring the shift of interference fringe were buried in the noise of low-sensitivity interferometers with air light's carriers (as shown by me in Fig.\ref{fig1}). In the case under consideration when the rays enter from the air into high optical density light-carrying medium by the wrong scheme shown in Fig.\ref{fig4}a, the interference pattern will be formed by short sections of the air arms of the interferometer $\Delta L_\bot$ and $\Delta L_\|$ whose length is $\sim 1-5$ cm.

  Taking into account that Michelson in 1881 with the length of interferometer arms $L_\bot=L_\|=1.2$ m observed a "null" shift amid the noise, as shown by curve 1 of Fig.\ref{fig1}, it is not difficult to understand that with the arms length $\sim 1-5$ cm the shift of the spurious interference fringe will be 100 times less. In other words, the intense spurious interference fringe will firmly stand still at any rotation of the interferometer, obscuring a much fainter principal interference fringe formed by the rays $S_{n\perp}$ and $S_{n\|}$ (seeing this in my experiment I have found the way to escape it). In the device with high density optical media there are many loci of arising a stray  interference which may lead to the methodical mistake of obtaining a "negative" result such as shown in  Fig.\ref{fig4}a.

In section 1 of this work, I discovered the secret of ''negativity'' of the experiments on Michelson interferometer with air light's carriers \cite{Demjanov rus, Demjanov}. By this publication I inform the scientific community about the methodical artifacts which may bury as ''negative'' all attempts of measuring on interferometers with high optical density light's carriers.  I experimentally found the described here effect observed by me in all cases where light's carriers were represented by cells (for gases or liquids) or rods with straight ends perpendicular to rays entering into them.

When placing in the way of the basic light beam in the Michelson interferometer rods or cells with optical media having the refraction index $n_c>n_{air}$ the light will be partially reflected  from the rectangular ends of the rods.   With the normal incidence of the beam on the media interface the coefficient of reflection is given by the expression
\begin{equation}
R=\left(\frac{n_{12}-1}{n_{12}+1}\right)^2,\label{reflection}
\end{equation}
where $n_{12}=n_c/n_{air}$. For the plexiglass $n_{12}=1.49$ we get $R=0.04$, and for the fused quartz $n_{12}=1.83$ we have $R=0.08$. Thus, for the rods with $1.5<n_{12}<2$ the part, from 4\% to 10\%, of orthogonal beams are mirrored back in air gaps of the length $\Delta L \sim 1-5$ cm to the splitting plate. These undesirable reflections form a short-arm spurious Michelson interferometer that, due to small scattering at short distances in the air, may ensure a very sharp interference pattern with the resolution of the fringe's relative shift $A_m=X_m/X_o\sim 10^{-3}-10^{-5}$ as in \cite{Shamir, Trimmer}.

By the estimations made the greater part of the light's energy, from 90\% to 96\%, enters the material of the rod. However, further the light is weakened in the visible part of the spectrum with the coefficient of attenuation from 1\% to 50\% at each cm $-$ even when propagating in such high-transparent media as plexiglass, fused quartz, various glasses.

In may experiments  the intensity of the light is weakened from 10 to 1000 times after passing there and back in rods of the length 5$-$100 cm made from the above mentioned stuff. This suggests from 0.1\% to 10\% of the original intensity. Thus, the ratio $\eta=S_n/S_r$ of the useful signal $S_n$ (that has passed to the rod and returned to the origin) to the intensity $S_r$ of the parasitic one (reflected immediately from the rectangular end of the rod and returned back to the splitting plate) is about 1. at best and 0.1 at worst. In other words, if the unwanted reflections from the ends are not removed, they clog and surpass the useful signal of the interference pattern obtained from of rays that passed through the rods and returned back.

In practise we see both interference patterns: one is more intensive (if reflections from the butts were not removed),  contrasting and "absolutely" immobile; another is less intensive, diffusive and markedly variable under the turning of interferometer by $90^\circ$.

Since the resolving power of the  Shamir\&Fox \cite{Shamir} setup has been much greater ($\sim 10^{-4}$)  than that of my installation  ($0.5\cdot10^{-2}$), the shift found by them was 1/3000, and I was not able to register at my device such a small quantity. So, I suppose that so small the shift found by Shamir\&Fox refers not to the whole length of the interferometer arms but to its small part $\Delta L_\perp=\Delta L_\|=1-5$ cm involving the air gap between the end of the rod and the splitting plate (see Fig.\ref{fig4}a). The lower resolution of my system appeared to be favorable in this case, since detecting no shift of the observed contrasting fringe but knowing from the previous experiments with gases and liquids that it must be significant I began to look for the cause of the failure, and have found it. Firstly, we may easily see that the contrasting spurious interference was not due to the rays $S_{n\perp}$ and $S_{n\|}$ that have passed through the plexiglas rods: the small angular rotation of the mirror or even interruption of the beam coming to the mirror does not affect the interference pattern. Secondly, a conventional placement of the rod leads to that the valid interference pattern is cluttered by more contrasting spurious reflections. To remove the harmful interference associated with reflections from the ends is feasible, for example, by small rotational displacement of the rod (by the angle $10-15^\circ$). In this case, due to that the incidence of rays on the end becomes oblique, the spurious rays are drawn aside, away from the interference screen (see Fig.\ref{fig4}). As a result, the contrasting interference pattern disappears and blurred interference pattern formed by the rods is brought to the fore. It is this interference pattern that proves to be susceptible to rotation of the interferometer in the horizontal plane. Processing by formula (\ref{aether wind Demjanov}) large relative shifts $A_m$ thus obtained with introducing the values of the dielectric permittivity gave the aether wind velocity that varied during the day and night in the range 140-480 km/s at the latitude of Obninsk, as shown in Fig.\ref{fig2}.

Fig.\ref{fig4} represents one of the configurations excluding zones of spurious interference. It enables us to single out the main motive of the interference obtained from the rays $S_{n\bot}$ and $S_{n\|}$ $-$ suppressing spurious rays $S_{r\bot}$  and $S_{r\|}$ by two radical ways:

$-$ oblique incidence of primary rays $S_\bot$ and $S_\|$ at appropriately angled butts of optical elements with sending reflected rays $S_{r\bot}$ and $S_{r\|}$ into absorbing traps $W_\bot$ and $W_\|$  (see Fig.\ref{fig4});

$-$ coating beveled butts of optical cells by special quarter-wave antireflection layers (arl), calculated for a given wavelength of light under the chosen angle of skewing and known refractive index.

In this way I was able to raise the signal/noise ratio $S_{n\bot}/S_{r\bot}$ and $S_{n\|}/S_{r\|}$ several hundred times performing measurements on the lengths of optical liquid and solid materials up to 1 m. However, the sensitivity of the Michelson interferometer in the range of permittivity $1.2<\varepsilon<3$ (excluding the region of $\varepsilon=2$) is so high that the length of cells or rods can be confined by values $5\leq L\leq 30$ cm. In this event the signal/noise ratio is no less than 10 (if to remove properly all the above described boundary effects) with the shift of the interference fringe being of the order of the very bandwidth. It is merely impossible not to discern such values of the shift.

\textbf{The second reason} of negative result may be concerned with the  phenomenon of complete loss of sensitivity of the interferometer to ''aether wind'' at  $\Delta\varepsilon = \varepsilon-1 = 1$ (see Fig.\ref{fig3}) \cite{Demjanov rus, Demjanov}. I observed this effect in the following media: water at a blue ray $\lambda=3\cdot10^{-7}$ m, plexiglass, polystyrene, transformer and capacitor oils, polyethylene at $\lambda=6\cdot10^{-7}$ m. The dielectric permittivity of these media lays in the range $1.95<\varepsilon<2.05$, where according to formula (\ref{shift Demjanov}) the shift of interference fringe changes sign and inevitably sinks in the noise of the device in the neighborhood of $\varepsilon=2$.

When trying to reproduce my measurements one need to exercise care and accuracy. So, in \cite{Shamir, Shamir Lipson, Cahill Kitto, Cahill} there is rather freely asserted that the refractive index of plexiglas is $n = 1.49$ (hence the  dielectric permittivity $\varepsilon=2.22$). Yet Shamir and Fox did not specify frequency of the measurement, which gave $n = 1.49$, and did not connect it with frequencies of rays in their installation. Over the past 40 years since the publication \cite{Shamir} almost all authors  referring to it uncritically replicated these values as some constants.

\begin{figure}[h]
  \begin{center}
\includegraphics[scale=0.6]{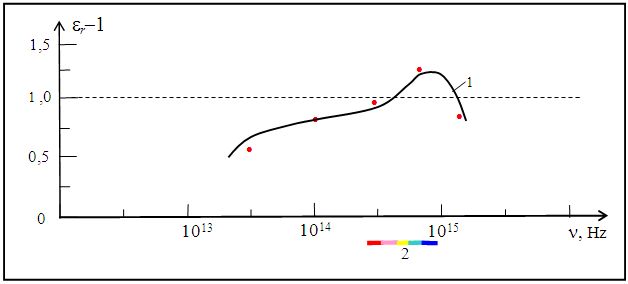}
  \caption{The dielectric permittivity $\varepsilon$ of the plexiglas as a function of the wave frequency $\nu$: the contribution $\varepsilon-1$ of particles is shown by curve 1. Measurements were made in the range of wavelengths from infrared to ultraviolet light by the method illustrated in Fig.\ref{fig4}a. The strip 2 shows the visible portion of the spectrum.}\label{fig5}
\end{center}
\end{figure}

In practise the dependence of $n$ on the wave frequency $\nu$ is essential. This dependence for plexiglas, measured by the method illustrated in Fig.\ref{fig4}b, is shown in Fig.\ref{fig5}.  You may see from curve 1 that value $n=1.49$ is observed somewhere in the green region of the spectrum, and it is different in other parts of it. In particular, for the white rays (with warm orange hue) of the incandescent lamp the plexiglas just has  $n\sim 1.4$, i.e. $\varepsilon\sim 2$. Therefore, the almost null shift of the interference fringe obtained in \cite{Shamir, Shamir Lipson, Cahill Kitto, Cahill} at plexiglas and presented without specification of the wave frequencies for the value $n=1.49$ and for the rays used in the interferometer of Shamir and Fox can not be accepted as a demonstration of the negativity of Michelson experiments at solid optical media. Rather these experiments corroborate the discovered by me dependence $X_m\sim\Delta\varepsilon(1-\Delta\varepsilon)$, according to which the  shift $X_m$ of the fringe passes through the zero at $n=1.41$, i.e. at $\Delta\varepsilon=\varepsilon-1=1$.

It remains for me to draw the attention of experimentalists to yet another simple scheme (it is shown in Fig.\ref{fig4}b) of preliminary testing optical-transparent materials concerning their suitability as light's carriers in Michelson interferometer. It is based on the use of oblique incidence (angle of incidence $\theta_i$) of a laser beam of selected wavelength on the horizontal plane of the vidicon's screen for fixing at it the point 1 of impact (initial test). Once installed on the path of the ray the cell (having parallel edges) with the gas or liquid, or  a solid sample (with parallel edges), as shown in Fig.\ref{fig4}b, by the shift ($b$) of the spot to point 2 (final test) there is determined the angle of refraction $\theta_n$ (hence, the refractive index $n$ or optical permittivity $\varepsilon=n^2$). By the ratio of the brightness of the spot after installing of the test material to the brightness of the spot before the installing there is assessed the optical attenuation (damping) per unit length of the beam in the medium. Errors from the cell sides are  accounted for trivially by methods of ray optics. These errors, of course, are reduced to the desired level by thinning-down the walls of the cell.

\section{The second means to determine velocity of "aether wind" $-$ via the winter relative to summer reduction (by 12$\%$) of the interference fringe shift}

The first way to determine the horizontal projection of the maximal velocity (designated by $V'$) of aether wind via the maximum amplitude  of the interference fringe shift (where $A_{m max}$ is measured in the course of the round-the-clock observation, as is shown in Fig.\ref{fig2}) consists in the calculation of $V'=V$ substituting $A_{m max}$ into formula (\ref{aether wind Demjanov}).

The second method became possible after that there were reliably measured the time dependencies $A_m(t_{\textrm{local}})$ of the fringe shift amplitude $A_m$ on the local time $t_{\textrm{local}}$ through the 24-hours cycle of the day and night (see Fig.\ref{fig2}). From dependencies $A_m(t_{\textrm{local}})$ in Fig.\ref{fig2} we may monitor the peak value  $A_{m\,\,max}$ through all seasons of the year. At the latitude of the city Obninsk the displacement of the $A_m(t_{\textrm{local}})$ peak appeared to be two hours per month (in local time scale). I found that during six months (from 22 December to 22 June) the peak of day-and-night dependence $A_m(t_{\textrm{local}})$ shifts by 12 hours and attains the maximum ratio $A_{m\,\,max\,\,max}(22.06)/A_{m\,\,max}(22.12)=1.12\pm 0.01$, i.e. it grows by $12\%$.

Independent astronomical observations of the projection of the sum of the Earth's linear velocity of motion around the Sun and Earth's linear velocity of motion around the center of Galactic shows that in summer (22 June) this sum equals to $\sim 205$ km/s and in winter (22 December) $\sim 235$ km/s \cite{Shklovsky}. Thus, if the seasonal increase (from December to June of the next year) of the maximal fringe shift in the horizontal plane of the device (by $\sim 12\%$) is due to change of the sum of the projection on the horizontal plane of the device of the Earth's round the Sun velocity and linear velocity of the Earth's motion around the center of Galactic, equaled to $\Delta V=235-205=30$ km/s, then the horizontal projection of the aether wind velocity at summer day-and-night peak (it is designated by $V''$) can be determined from the ratio of summer $V''$ and winter $V''-\Delta V$ velocities of the interferometer obtained according to (\ref{aether wind Demjanov}):
\begin{equation}
\frac{V''}{V''-\Delta V}=\left(\frac{A_{m\,\,max\,\,max}}{A_{m\,\,max}}\right)^{1/2}=\sqrt{\xi}.\label{peak displacement}
\end{equation}
By (\ref{peak displacement}) the peak summer velocity of the Earth relative to aether equals to $V''=\Delta V/(\sqrt{\xi}-1)=30/0.06\sim 500$ km/s. Clearly, the indirect estimating of $V''$ by the second means, from the maximal shift of the interference fringe in 24-hour cycle of the observation of the dependence $A_m(t_{\textrm{local}})$ in the areas of summer ($A_{m\,\,max\,\,max}$) and winter ($A_{m\,\,max}$) day-and-night peaks (Fig.\ref{fig2}) is well agreed ($V''\approx V'$) with the direct measurement of the peak aether wind velocity ($V'=480$ km/s) performed in the day-and-night cycle ($140-480$ km/s) \cite{Demjanov rus, Demjanov}.

\section{Experimental means to protect the interference pattern against its distortion by reflections from the light spot excited in the semitransparent layer by the primary ray}

\begin{figure}[h]
  \begin{center}
\includegraphics[scale=0.6]{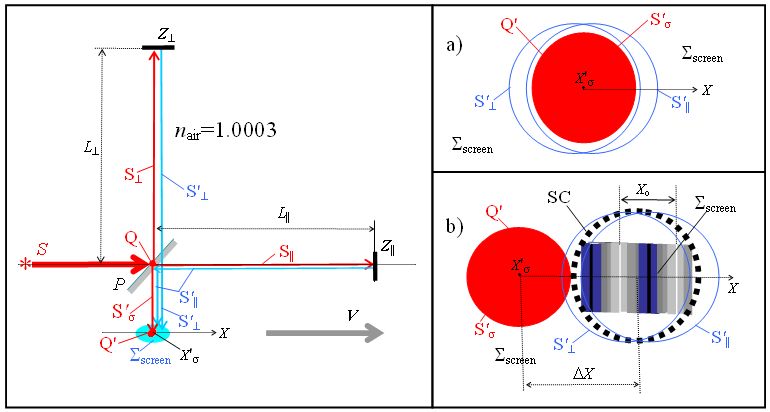}
  \caption{Optical scheme of splitting the ray from the source $S$ at the semitransparent plate $P$ of  the Michelson interferometer into two operating orthogonal rays $S_\|$ and $S_\perp$, which propagate to mirrors  $Z_\|$ and $Z_\perp$, and a spurious ray $S'_\sigma$, scattered to the interference screen $\Sigma_{\textrm{screen}}$, where the first two rays are interfered. There are shown in the insets:
a) the point of bifurcation $Q$ and two approximately brought together in it the returned rays $S'_\|$ and $S'_\perp$; this explains the "triple interference" at the plane of the screen of these beams with the third one $S'_\sigma$ that is the field of the spurious scattering of the primary beam $S$ from the source in the direction of the interference screen (here the resolving power by the shift of the fringe does not exceed 1/10 of the fringe width $X_\textrm{o}$);
   b) the point of bifurcation $Q$ and shifted interference fringe $X_\textrm{o}$ at the interference screen $\Sigma_{\textrm{screen}}$. The desired displacement of the zone of interference of operating rays $S'_\|$ and $S'_\perp$ at the plane of the interference screen away from the zone $S'_\sigma$ of maximal luminosity of the spurious scattering of the beam $S$ enhances the resolving power by the shift of the fringe up to 1/50 of the fringe  width $X_\textrm{o}$; using the cylinder-shaped shield (SC) enables us to increase it up to 1/200.}\label{fig6}
\end{center}
\end{figure}

There were described in section 4 two effects of non-observability of the interference fringe shift in the interferometer of Michelson type. The instrumental effect is concerned with the parasitic reflection of rays from the interface of the optical media with the refractive index $n>n_{\textrm{air}}$ placed in the arms of the interferometer. The physical one is connected with phenomenon of the full loss of the sensitivity of the interferometer to shift of the interference fringe when the refractive index of the both optical media becomes $n=\sqrt{2}$ (see Fig.\ref{fig2}). I will consider below another instrumental effect that is concerned with a parasitic scattering of the primary ray in the point $Q$ of bifurcation located at the semitransparent light-splitting plate (see Fig.\ref{fig6}).

Fig.\ref{fig6} illustrates one of the hidden causes of spurious interference in the Michelson  interferometer. Shown in this figure the optical scheme explains the existence of not only the bifurcation by the splitting plate of the main beam arrived from the source $S$, but the formation of the third leakage flux from the light spot $Q$. The third light flux $S'_\sigma$ is spuriously scattered from the zone $Q$ of the incidence of the primary beam, coming from the light source $S$, into the direction of interference screen $\Sigma_\textrm{screen}$ (i.e. in the opposite direction to the ray $S_\perp$), penetrating through a translucent layer of the splitting plate $P$. Scattered (0.1$\%$) in the direction of the screen $\Sigma_{\textrm{screen}}$ the light intensity $S'_\sigma$ seems relatively small ($<S/1000$) in circumstances when 0.999 of the entire source intensity goes on the formation of two orthogonal operating beams. However, the intensity of light in the returning operating rays $S'_\|$ and $S'_\perp$ is greatly attenuated (from 500 to 10000 times, depending on the length of the interferometer arms, and losses in the body of the optical medium). If the effect of spuriously scattered radiation is not suppressed (such as shown in Fig.\ref{fig6}b), when there come on the interference screen three coherent light fluxes of comparable intensity (as shown in Fig.\ref{fig6}a) which interfere with each other. Of the three pairs of interference patterns $S'_\|\leftrightarrow S'_\perp$, $S'_\|\rightarrow S'_\sigma$ and $S'_\perp\rightarrow S'_\sigma$ only the first is an objective characteristic of the sought-for speed of "ether wind", while the other two make one or other contributions to the systematic experimental error. If the rods or cells with optical media are installed in the arms of the interferometer then these three patterns are added with seven more ones! In the result, the resolving power with relation to the shift of the interference shift drops sharply $-$ up to a total loss of observability of the sought-for shift of the fringe $S'_\|\leftrightarrow S'_\perp$, produced when the interferometer is rotated in the horizontal plane.

\begin{figure}[h]
  \begin{center}
\includegraphics[scale=0.6]{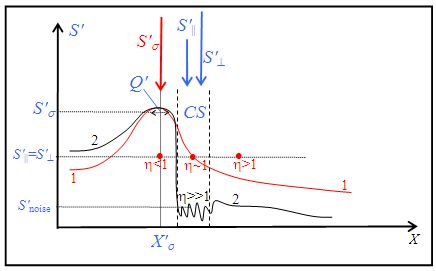}
  \caption{The intensity $S'_\sigma(X)$ of the light scattered from the bifurcation spot of the area Q on the semitransparent layer of the splitting plate $P$ as a function of the coordinate $X$ on the interference screen:
1 $-$ without a cylindrical shield (CS), when in the case of passing $S'_\|$ and $S'_\perp$ through the center of the spot $Q'$ bad  ratios signal/noise $\eta=S'_\|/S'_\sigma<1$ and $\eta=S'_\perp/S'_\sigma<1$ are raised merely by a displacement of the interference fringe obtained from the rays $S'\perp\leftrightarrow S'_\|$ to the level $\sim 1$;
2 $-$ with displaced away from the center of the spot $Q'$ a cylindrical shield (CS) having the inner diameter 3 mm and height 2 cm and mounted on the interference screen $\Sigma_{screen}$; in this case the signal/noise ratios become $\eta=S'_\|/S'_\textrm{noise}>>1$ and $\eta=S'_\perp/S'_\textrm{noise}>>1$.}\label{fig7}
\end{center}
\end{figure}

Choosing the region of interference of the two major orthogonal rays $S'_\|$, and $S'_\perp$ away from the strongly luminous point $Q$ (as shown in Fig.\ref{fig6}b and Fig.\ref{fig7}) can reduce the disturbances from above described spurious interference and increase the resolving power in relation to the shift of the fringe  up to 1/50. The means of optimization of the experiment by reducing this kind of the optical noise was described nowhere until now, so they are considered here in such detail. However we must mention that Michelson$\&$Morley with $L_\|=L_\perp =11$ m obtained the resolution by the shift of the fringe $\sim 1/40$ \cite{Michelson Morley},  and Miller with $L_\|=L_\perp =32$ m achieved $\sim 1/30$ \cite{Miller}, ], that I was not able to achieve (without displacement the brought together rays by $\Delta X\sim 5$ mm away from the luminous bifurcation point $Q'$, see Fig.\ref{fig6}) even for smaller arms $L_\|=L_\perp=7$ m in the air. Therefore, we may conjecture that they displaced (even may be by accident and empirically)  the zone of interference of the two main beams $S'_\|$, and $S'_\perp$ away from the luminous zone $Q$ on the plate $P$ as shown in Fig.\ref{fig6}b and Fig.\ref{fig7}.

 The means to increase the resolving power of the classical Michelson interferometer, such as to displace the interference pattern from the zone $Q$, can be significantly improved. To do this in my experiments I installed the shield cylinder (SC) made of graphite, which separate a couple of useful beams $S'_\|$ and $S'_\perp$ from stray reflections. The inner diameter of the cylinder was chosen somewhat larger than diameters of the rays $S'_\|$ and $S'_\perp$ (see Fig.\ref{fig6}b). I used the cylindrical shield of the height $2$ cm and inner diameter $3$ mm, which was mounted on the interference screen with the displacement of its axis from the point $Q$ of bifurcation by $5-7$ mm. Thus the residual noise $S'_\textrm{noise}$ was reduced several times more comparing with curve 1 in Fig.\ref{fig7}. In this way I was able to increase the resolving power of the  interferometer with length of rays in the air $L_\|= L_\perp = 7$ m  to 1/200 of the width of the fringe. For fused quartz and water and with the above described means to reduce spurious reflections using beveled (at $\sim 15^{\circ}$) ends even with the arms length $L_\|=L_\perp=60$ cm I was able to obtain the resolution up to $\sim 1/100$ of the fringe width (as described in \cite{Demjanov rus, Demjanov}).

Below there are few words concerning the means how to configure the optical system in order to obtain on the principal  interference fringe the minimum noise arising from spurious reflections emanating from the bifurcation point $Q$  of the source's light beam. The technique involves three main regulations:

1) The beam $S'_\|$ is removed installing in the way of the beam $S_\|$ (right before the mirror $Z_\|$) a plate absorbing the light. Thus the uniform illumination by the beam $S'_\perp$ reflected from the mirror $Z_\perp$ is attained at the interference screen $\Sigma_{screen}$. To this end by small rotation of the mirror $Z_\perp$ such a displacement of the ray $S'_\perp$ from the point $Q$ of bifurcation of the source's light is attained that a scent of spurious interference of rays $S'_\perp\leftrightarrow S'_\sigma$ entirely disappears.

2) Installing in the way of the beam $S_\perp$ (right before the mirror $Z_\perp$) a plate absorbing the light removes  the beam $S'_\perp$. Thus the uniform illumination by the beam $S'_\|$ reflected from the mirror $Z_\|$ is attained at the interference screen $\Sigma_{screen}$. In this procedure the displacement of the beam $S'_\|$ from the point $Q$ of bifurcation of the light beam arrived from the source is carried out more accurately. Thus the signs of the spurious interference $S'_\|\leftrightarrow S'_\sigma$ in this channel entirely disappear; procedures described in paragraphs 1 and 2  are repeated iteratively until the spurious interference will be fully eliminated.

3) The beams $S_\perp$ and $S'_\|$ are eliminated installing in their way the light-absorbing plates (as in paragraphs 1 and 2). Then there must be observed at the interference screen the "null" illumination $S_{\textrm{noise}}$ due to residual light noise of the device. In this regulation experiments I managed to obtain the signal-to-noise ration ($S'_\|/S_{\textrm{noise}}\sim S'_\perp/S_{\textrm{noise}}$) up to $10-100$ for water, plexiglas and fused quartz.

After the above adjustment of the optical system, we open the passage of both beams to the mirrors $Z_\perp$ and $Z_\|$. As a result, we obtain the sought for interference fringe pattern from two main beams $S'_\|$ and $S'_\perp$. It is this pattern that shifts when the interferometer is turned in the horizontal plane as shown in Figures \ref{fig2} and \ref{fig3} and described in \cite{Demjanov rus, Demjanov}. The spurious interference patterns, eliminated by the above described method,  remain actually invariable when you turn the interferometer.

Let us estimate the extent of the fringe shift in an interferometer with the rods of glass or fused quartz. With the length of the rods $L_\perp= L_\|\sim 10$ cm in summer 20 June at noon or at midnight in winter 20 December at the latitude of Obninsk ($\sim 56^\circ$ N) the maximum shift of the fringe reaches the width of the interference fringe (see Fig.\ref{fig1} curve 4 and Fig.\ref{fig3} point 6). Such a shift is impossible to be unnoticed!

\section{Collisions in interpreting processes occurring in Michelson interferometer}

In the current section we will show that the interpretation of the expected results in Michelson experiments could not be successful in the 19th century being based on first principles of the classical (Galilean) theory of relativity. From the very beginning these experiments required the knowledge of nonclassical relativistic Lorentz-invariance of the light's speed not only in vacuum, but also in the absolute motion of optical media with the refraction index $n>1$.

\textbf{7.1. Period (1880-1960) of disregarding the decisive role of the medium with} $\bf{n>1}$ \textbf{in the interpretation of Michelson experiments.} In 1881 Michelson started from the geometry-optical model  of the experiment whose idea was put forward by Maxwell. Michelson believed still before the experiment that the difference of times that the light propagates in orthogonal arms of the device should be nonvanishing. Making calculations for the speed $c/n$ of light in vacuum medium ($n=1$), he determined the time that light propagates in the arm $L_\|$  parallel to $\mathbf{V}$ for the case $V_\|=|\mathbf{V}|$ in vacuum when $c/n= c/1.=c$:
\begin{equation}
t_\|=\frac{L_\|}{c-V}+\frac{L_\|}{c+V}=\frac{2L_\|}{c}\frac{1}{1-V^2/c^2}.\label{time parallel Michelson}
\end{equation}
Similarly, in the arm $L_\perp$  perpendicular to $\mathbf{V}$, in which with $V_\|=|\mathbf{V}|$ the projection $V_\perp=0$, in order to determine the light propagation time he used the same logic as in (\ref{time parallel Michelson}):
 \begin{equation}
t_\perp=\frac{2L_\perp}{c}.\label{time perpendicular Michelson}
\end{equation}
In the outcome he obtained the following nonzero difference of times for $n=1$ and $L_\|=L_\perp =L$:
\begin{equation}
\Delta t = t_\perp - t_\| \approx  - \frac{2L}{c}\frac{V^2}{c^2}.\label{time difference Michelson}
\end{equation}
Substituting  (\ref{time difference Michelson}) into the expression of the transverse (to the beams) relative shift $A_m$ of the interference fringe (${A_m} = c\Delta t/\lambda $ ), Michelson by the orbital speed of the Earth about the Sun ($\sim 30$ km/s) estimated the expected from (\ref{shift}) amplitude ($A_{m\,\,\mathtt{exp}.}$) of the fringe shift:  $A_{m\,\,\mathtt{exp}.}=0.04$ \cite{Michelson}. In 20th century became clear that the speed of the Earth in  space is really not less than 400 km/s \cite{Shklovsky}.

Having in 1881 the experiment carried out for the arms length $L=1.2$ m \cite{Michelson}, he obtained the "negative" (zero: $A_m=0$) result, whereas the positive result was expected to be obvious for the device resolution $A_m\sim 0.025$. Scientists began to look for causes of these overestimations. The first who found two-fold overestimations in (\ref{shift}) was Lorentz. After correction (\ref{time parallel Michelson}-\ref{time difference Michelson}) by the so called "Lorentz triangle" formula (\ref{time perpendicular Michelson}) became:
\begin{equation}
{t_ \bot } = \frac{{2{L_ \bot }}}{c}\frac{1}{{\sqrt {1 - {V^2}/{c^2}} }}.\label{time perpendicular vacuum}
\end{equation}
Accounting for (\ref{time perpendicular vacuum}) in (\ref{time difference Michelson}) gave two times smaller value of  $\Delta t$, than it was in the Michelson calculation \cite{Michelson} (but as before nonzero when $n=1$):
\begin{equation}
\Delta t = {t_ \bot } - {t_\parallel } = \frac{{2L}}{c}\left[ {\frac{1}{{\sqrt {1 - {V^2}/{c^2}} }} - \frac{1}{{1 - {V^2}/{c^2}}}} \right] \approx  - \frac{L}{c}\frac{{{V^2}}}{{{c^2}}}.\label{time difference Lorentz triangle}
\end{equation}

In 1887 Michelson$\&$Morley repeated the experiment \cite{Michelson Morley}, increasing by the order of magnitude the arm's length (up to $L=11$ m). They expected to obtain 10 times greater shift amplitude [from (\ref{time difference Lorentz triangle}) it should be no less than $A_{m\,\,\mathtt{exp.}}=0.2$]. Obviously, at the device resolution power  $\delta A_m\sim 0.035$ with the arm's length 11 m it is impossible not to notice the fringe shift $A_m\sim 0.2$. However, at $L=11$ m Michelson and Morley did not find too the expected shift of the interference shift (curve 2 in Fig.3). The theoretical search for explaining this phenomenon got under way.

Now already with the knowledge of the mathematical structure (\ref{time difference Lorentz triangle}), corrected by the "Lorentz triangle", Fitzgerald and independently Lorentz suggested that the length $L_\|$ of the arm parallel to $\mathbf{V}$ is shortened dynamically (this phenomenon was named "Lorentz contraction") as  $L{'_\parallel } = {L_\parallel }\sqrt {1 - {V^2}/{c^2}} $. Accounting for $L{'_\parallel }$  in (\ref{time difference Lorentz triangle}) indeed gave the zero "theoretical" shift of the fringe:
\begin{equation}
\Delta t = {t_ \bot } - {t_\parallel } = \frac{{2L}}{c}\left[ {\frac{1}{{\sqrt {1 - {V^2}/{c^2}} }} - \frac{{\sqrt {1 - {V^2}/{c^2}} }}{{1 - {V^2}/{c^2}}}} \right] = 0.\label{time difference vacuum}
\end{equation}
Substantially, that only with two relativistic corrections of the original classical form (\ref{time difference Michelson}) of Michelson by the "Lorentz relativistic radicals"   we arrive at the result (\ref{time difference vacuum}). They are: the "Lorentz triangle" in the arm ${L_ \bot }$  and Lorentz contraction in the arm  ${L_\parallel }$. Only after these two relativistic corrections in (\ref{time difference vacuum}) we obtain zero and solely for $n=1$. No other corrections give null in the original classical Michelson formula (\ref{time difference Michelson}) at $n=1$.

In 1920s Miller, having increased the arm's length of the interferometer up to $L=32$ m, raised three times the expected from (\ref{shift}) magnitude of the fringe shift \cite{Miller} (in comparison with $A_{m\,\,\mathtt{exp}.}\sim 0.4$ in \cite{Michelson Morley}), bringing it to $A_{m\,\,\mathtt{exp}.}\sim 1.2$. Owing to this, Miller for the first time confidently registered the fringe shift at the ratio signal/noise=2 (see curve 3 in Fig.1). However, the value $A_m$ measured by him appeared $\sim 10$ times less than that expected by (\ref{shift}). Besides, the fringe shift has been reliably observed only during 1-3 hours and disappeared in other time of day and night (see Fig.1, curve3). When the unstable as they are results of Miller have been verified by Joos in vacuum \cite{Joos}, no signs of the fringe shift were found (the period of the ignorance of the decisive role of the medium with $n>1$ for experiments of Michelson type continued). Since then the majority of scientists became to consider Michelson type experiments "negative", taking no notice until the middle of the 20th century of the fact that the relationship (\ref{time difference vacuum}) is valid solely in vacuum (when $n=1$). Only in 1960s there was realized \cite{Demjanov rus, Demjanov, Shamir, Cahill} that the measurements made in vacuum always must give the zero difference  $\Delta t = {t_ \bot } - {t_\parallel }$ of times that the light propagates in orthogonal arms of the interferometer. But until now many believe the Lorentz contraction to be a mathematical trick "devised" in order to explain the "null result" (\ref{time difference vacuum}) of experiments in all media. The investigations \cite{Demjanov rus, Demjanov} have showed that such an extension of (\ref{time difference vacuum}) is mistaken.

\textbf{7.2. The period (1960-2010) of awakening the attention to the decisive role of the medium with} $\mathbf{n>1}$ \textbf{in the interpretation of experiments at interferometers of Michelson type.} In 1968-74 years I reproduced Michelson$\&$Morley and Miller experiments not only in a weak laboratory vacuum of the density $0.01-0.5$ atm, but also made them in the air of the density $1-3$ atm and in various optically transparent gases (see Fig.3). Then the linear regions of the curves 1 and 2 in Fig.3 were revealed. For the first time there was followed from experiments the necessity that  polarizing particles to be present in the light carrying medium of the interferometer in order that the detecting medium of the light carriers has $n>1$ (i.e. has the dielectric permittivity $\varepsilon =n^2=(1.+ \Delta\varepsilon) >1$, where $\Delta\varepsilon >0$ is the contribution of particles into the permittivity of the optical medium).

The sensitivity of the interferometer to the shift of the interference fringe appeared to grow proportionally to the contribution $\Delta\varepsilon=n^2-1$ of the particles' polarization into the permittivity of the light carriers, and the noise of the device grows proportionally to $\varepsilon$. The Fresnel ratio $(1-n^{-2})=\Delta\varepsilon/\varepsilon$,  known to this time already $\sim 150$ years, appeared to be a key parameter determining the sensitivity of the interferometer. The amplitude of the fringe shift $A_m(\varepsilon,\Delta\varepsilon, \Delta\varepsilon^2)$ proved to be nonlinear dependent just on the contribution    of the particles of the light carrier (\ref{shift Demjanov}), turning to null at  $\varepsilon=1$ and  $\varepsilon=2$ (see Fig.3).

This new set of positive (non-zero) results needed an explanation on the basis of the modern knowledge. Nowadays it is known a few such explanations. The experimental results described in \cite{Demjanov rus, Demjanov} (in the whole they are represented above in Fig.3) I interpreted by means of corresponding corrections to Michelson model (\ref{time parallel Michelson}-\ref{time difference Michelson}). Lorentz and Fitzgerald in the end of 19th century were guided by  results \cite{Michelson, Michelson Morley} above shown in Fig.1 by curves 1 and 2 which interpreted by them as a "proof of the absence" of the fringe shift ($\Delta t=0$ and $A_m=0$). The experimental guide for me since 1970 became the obtained by me dependencies $A_m(\varepsilon,\Delta\varepsilon, \Delta\varepsilon^2)\neq0$ given in Fig.2 and 3 where the fringe shift is absent ($A_m=0$) only at two values of the medium polarization: $\Delta\varepsilon=0$, $n=1$; and  $\Delta\varepsilon=1$, $n=1.41$, $\varepsilon=2$. Being guided by my experimental results in 1971 I have found the theoretical dependence (\ref{shift Demjanov}).

In order to do this, in the arm parallel to $\mathbf{V}$ instead of $\tilde{c}  =c  \pm V$ valid for vacuum ($n=1$) I used Fresnel formula  $\tilde{c} = c/n \pm V(1 - {n^{ - 2}})$ accounting for the polarization of particles of light carriers by the refractive index $n>1$. Further I introduced in the longitudinal arm the Lorentz time dilation  $t{'_\parallel } = {t_\parallel }/\sqrt {1 - {V^2}/{c^2}} $, and in the arm perpendicular to $\mathbf{V}$  I have computed the "Lorentz triangle " not in vacuum ($n=1$) as Lorentz did but in an optical medium with  $n>1$. Introduced by me in this way relativistic corrections of the expressions $t{'_\parallel }$ and  $t{'_ \bot }$ can be supposed to make invariant the whole calculation of $\Delta t$ based on the prerelativistic Fresnel formula. In the outcome they yielded the appropriate (for description of Fig.3) formula (\ref{shift Demjanov}).

After the publication \cite{Demjanov} authors \cite{Dmitriyev, Morris, Demjanov first order} suggested three different derivations of the formula (\ref{shift Demjanov}). There was shown in \cite{Dmitriyev} that the noninvariance of the Fresnel formula in the arm parallel to $\mathbf{V}$ can be corrected otherwise than it was in \cite{Demjanov rus, Demjanov}, namely by means of the relativistic correction of the time dilation in the transverse arm, and then formula (\ref{shift Demjanov}) get obtained as well. There was also shown in \cite{Dmitriyev} that the similar result can be achieved by another approach to construction the analog of "Lorentz triangle" for the transverse interferometer arm. It should be stressed that the description of the experimental dependence $A_m(\varepsilon,\Delta\varepsilon)$ in Fig.3 in all cases listed was attained by the introducing into the classical schemes of the calculation the corrections by that or another form of the relativistic "Lorentz factor" $\sqrt{1 - {V^2}/{c^2}}$.

An original derivation of the formula (\ref{shift Demjanov}), not intersecting with the Michelson and Lorentz model, was proposed by the author \cite{Morris}. He introduced the relativistic correction $(1/\sqrt {1 - {V^2}/{c^2}})^2$ into the dispersion relation by Maxwell-Zellmeier for the refractive index and obtained anisotropic expressions for ${n_\parallel }$  of the medium in the arm parallel to $\mathbf{V}$ and  ${n_ \bot }$ of the medium in the arm perpendicular to $\mathbf{V}$. Using these expressions for determination of  $t{'_\parallel } = {L_\parallel }/(c/{n_\parallel })$  and $t{'_ \bot } = {L_ \bot }/(c/{n_ \bot })$  he obtained formula (\ref{shift Demjanov}). In \cite{Demjanov first order} I have developed the idea of the work \cite{Morris} and demonstrated that if instead of the relativistic correction $(1/\sqrt {1 - {V^2}/{c^2}})^2$ suggested in \cite{Morris} we represent the wave length in the Maxwell-Zellmeier dispersion relation by the modern relativistic Lorentz-invariant expression of the Doppler coefficient of the first and second order by $V/c$
\begin{equation}
D_{ + / - }^2 = \frac{{{{(1 + {V \mathord{\left/
 {\vphantom {V c}} \right.
 \kern-\nulldelimiterspace} c} \cdot \cos {\theta _{0/180}})}^2}}}{{1 - {{{{{V^2}} \mathord{\left/
 {\vphantom {{{V^2}} c}} \right.
 \kern-\nulldelimiterspace} c}}^2}}} \simeq (1{\raise0.7ex\hbox{$ + $} \!\mathord{\left/
 {\vphantom { +   - }}\right.\kern-\nulldelimiterspace}
\!\lower0.7ex\hbox{$ - $}}\,2{V \mathord{\left/
 {\vphantom {V c}} \right.
 \kern-\nulldelimiterspace} c} + 2{{{{V^2}} \mathord{\left/
 {\vphantom {{{V^2}} c}} \right.
 \kern-\nulldelimiterspace} c}^2}),\label{Doppler}
\end{equation}
a new generalization of the formula (\ref{shift Demjanov}) can be obtained which describes the shift of the interference fringes in the interferometers capable in special circuit realizations \cite{Demjanov first order} to work at effects of first and second order by $V/c$. Generalized in this way the expression of the time difference for propagation of light in perpendicular and parallel arms becomes:
\begin{equation}
\Delta t \approx \frac{L}{c}\left[ {\frac{V}{c}\left( {\frac{\Delta {\varepsilon _1}(1 - \Delta \varepsilon _1)}{{{n_1}}} - \frac{\Delta {\varepsilon _2}(1 - \Delta \varepsilon _2)}{{{n_2}}}} \right) + \frac{{{V^2}}}{{{c^2}}}\left( {\frac{\Delta {\varepsilon _1}(1 - \Delta \varepsilon _1)}{{{n_1}}} + \frac{\Delta {\varepsilon _2}(1 - \Delta \varepsilon _2)}{{{n_2}}}} \right)} \right].\label{time difference Doppler}
\end{equation}
Substituting (\ref{time difference Doppler}) into the expression of the amplitude $A_m=c\Delta t/\lambda$  of the fringe shift depending on $\Delta t$ gives the following generalization of (\ref{shift Demjanov}):
\begin{equation}
{A_m} \approx \frac{L}{\lambda }\left[ {\frac{V}{c}\left( {\frac{\Delta {\varepsilon _1}(1 - \Delta \varepsilon _1)}{{{n_1}}} - \frac{\Delta {\varepsilon _2}(1 - \Delta \varepsilon _2)}{{{n_2}}}} \right) + \frac{{{V^2}}}{{{c^2}}}\left( {\frac{\Delta {\varepsilon _1}(1 - \Delta \varepsilon _1)}{{{n_1}}} + \frac{\Delta {\varepsilon _2}(1 - \Delta \varepsilon _2)}{{{n_2}}}} \right)} \right]\label{shift Doppler}
\end{equation}
which when $n_1=n_2$ (i.e. $\Delta {\varepsilon _1} = \Delta {\varepsilon _2}$), due to obvious from (\ref{shift Doppler}) compensation of the first order by $V/c$  effects, as it takes place in the classical Michelson interferometer, turns into (\ref{shift Demjanov}). The successful experimental testing of the possibility to construct Michelson interferometer operating at first order by $V/c$ effects has been performed and described in \cite{Demjanov rus} and \cite{Demjanov first order}.

Thus, the accumulated experience of the theoretical description of the kinetic phenomenon of aether wind observed by means of Michelson interferometer testify to necessity of introducing relativistic corrections to the classical scheme of description of velocity characteristics of the light's propagation in moving medium. These corrections seem to make invariant that or another part of the classical mathematical model used to calculate the shift of the interference fringe ensuring its agreement with the experiment (see Fig.\ref{fig3}).

I will show now below that if from the very beginning to use the known to date relativistically invariant mathematical tools describing the propagation of light in moving media then formula (\ref{shift Demjanov}) arises in a natural way, without any other corrections.

\textbf{7.3. The accounting for polarization of particles of optical media, with the simultaneous using the relativistically invariant tools, necessary in order to describe and interpret results of the Michelson-type experiments.} As can be seen from the classical scheme (\ref{time parallel Michelson}) of calculating the time of propagation of light in an optical medium, we should know a formula for the speed of propagation. Michelson did not know (was not able to know in 1881) all particulars of the object under investigation, in the first place, that for processing the results of non-zero measurements $A_m\neq 0$ a nonclassical definition of the light propagation velocity   will be needed. Michelson could not know that the light speed should be a non-Galilean invariant of the moving and stationary inertial reference frames related accordingly with stationary aether and inertial particles of the light carrying medium moving in aether.

Secondly, there was not known either a constructive approach to solution of the problem posed demanding optical media with $n>1$ in order to attain the observation of $A_m\neq 0$, and for correct interpreting the result of measurements of the fringe shift $-$ the obligatory taking into account $\varepsilon=n^2$  (for the air $n\approx 1.0003$ with the accuracy up to 4th digit after 1.). An invariant formula of the light's speed to describe dynamics of these two inertial frames of reference can be obtained  providing to use the many times verified in 20th century in experiments with moving media formula of relativistic rule for addition of velocities in vacuum:
\begin{equation}
\tilde{c} = c \oplus V = \frac{{c \pm V}}{{1 \pm cV/{c^2}}} = \frac{{c \pm V}}{{1 \pm V/c}}=c.\label{addition vacuum}
\end{equation}
In this formula $c=c/n=c/1.$ is the speed of light in the medium without particles (i.e. in the stationary aether where $n=1.$), and operator $\oplus$  means the rule of relativistic addition, and $V$ is the translational velocity that  particles of light carrier move in aether. Formula (\ref{addition vacuum}), in its narrow content,  was discovered only in 1904 by Poincare (in particular for calculating the phenomenon of drag of the light complex by moving particles, but not vice versa as thought earlier):

The sign "+" corresponds to coincidence of the direction of vectors $\mathbf{c}$ and $\mathbf{V}$, and sign "$-$" to their contraposition. This scheme follows from the known Lorentz group transformations incorporating used above in the capacity of corrections the Lorentz factor $\sqrt {1 - {V^2}/{c^2}} $.

The widely known formula (\ref{addition vacuum}), underlying the definition and proof of a part of the second postulate of special relativity, is a particular case of a more general formula which is unfortunately less known:
\begin{equation}
\tilde{c}_\pm = \,\,\,c/n \oplus V\,\,\, = \,\,\,\frac{{c/n \pm V}}{{1 \pm \dfrac{{Vc/n}}{{{c^2}}}}}\,\,\,\, = \,\,\,\,\dfrac{{c/n \pm V}}{{1 \pm V/(nc)}}.\label{addition medium}
\end{equation}
When $n=1$  formula (\ref{addition medium}) turns into (\ref{addition vacuum}). Similar to formula (\ref{addition vacuum}) in relation to unitary invariant $c$, the formula (\ref{addition medium}) is invariant in relation to combination $c/n$. It can be easily shown. If the direct (for variation $\delta c=\pm V$) transformation of velocity $c/n$  of light in the stationary medium using (\ref{addition medium}) yields $\tilde{c}_\pm$, then the reverse transformation of $\tilde{c}_\pm$ (for variation $\delta c=\mp V$)  gives:
\begin{equation}
\tilde{c}' = \tilde{c}_\pm \oplus V = \frac{{(\tilde{c}_\pm \mp V)}}{{1 \mp \tilde{c}_\pm V/{c^2}}} = c/n,\label{addition reverse}
\end{equation}
i.e. again the same velocity speed in the stationary medium. Let us find by the above described way the expression of speed of light invariant both in moving and stationary media. Thereto expand the right hand part of (\ref{addition medium}) into series by a small parameter $\pm V/c$:
\begin{equation}
\tilde{ c}_\pm \approx \frac{c}{n}\left[ {(1 \pm \frac{V}{c}n\left( {1 - \frac{1}{{{n^2}}}} \right) - \frac{{{V^2}}}{{{c^2}}}\left( {1 - \frac{1}{{{n^2}}}} \right) \pm \frac{{{V^3}}}{{{c^3}}}\frac{1}{n}\left( {1 - \frac{1}{{{n^2}}}} \right) - \frac{{{V^4}}}{{{c^4}}}\frac{1}{{{n^2}}}\left( {1 - \frac{1}{{{n^2}}}} \right) \pm ....} \right].\label{expansion}
\end{equation}
The first two terms of this series is known to give the prerelativistic formula by Fresnel $\tilde{c}_\pm = c/n \pm V(1 - {n^{ - 2}})$ . This result is ascribed to one of the first achievements of special relativity produced a phenomenological derivation of this formula \cite{Pauli}. There was shown in \cite{Dmitriyev} that this formula $\tilde{c}_\pm = c/n \pm V(1 - {n^{ - 2}})$  is noninvariant with respect to the combination $c/n$.

It appeared that if the direct transformation of velocity by (\ref{expansion}) is abridged not by two but three terms of the series incorporating the first and second order of the ratio $V/c$
\begin{equation}
\tilde{c}_\pm\approx \left[ {\frac{c}{n} \pm V\left( {1 - \frac{1}{{{n^2}}}} \right) - \frac{{{V^2}}}{{cn}}\left( {1 - \frac{1}{{{n^2}}}} \right)} \right],\label{expansion abridged}
\end{equation}
then performing the reciprocal transformation by means of (\ref{addition reverse}) with the use of (\ref{expansion abridged}) gives again the same result $\tilde{c}'=c/n$. This indicated that the refined (by retaining the terms of the second order $V/c$) Fresnel formula (\ref{expansion abridged}) becomes invariant with respect to parameter $c/n$  both for moving and stationary media (taking into account only the effects of the first and second order in $V/c$). Here we encounter with interesting mathematical fact: using in the transformation (\ref{addition reverse}) both the exact form (\ref{addition medium}) for $\tilde{c}_\pm$ and approximate form (\ref{expansion abridged}) for $\tilde{c}_\pm$ , in which discarded all the terms in the expansion (\ref{addition medium}) of the order magnitude higher than $O(V^3/c^3…)$, in both cases yields by (\ref{addition reverse}) the same invariant result $\tilde{c}'=c/n$ (that is invariant for moving and stationary media or associated with them inertial reference frames). One must only remember, that (\ref{expansion abridged}) is valid solely for describing processes in the absolute (stationary) and laboratory (moving) inertial reference frames to interpret effects taking into account first and second degree of smallness in the ratio $V/c$.

In this approach, the lag in the arm parallel to $\mathbf{V}$, which is computed using Lorentz-invariant formula (\ref{addition reverse}) in (\ref{time parallel Michelson}), yields at once the structure of formula (\ref{shift Demjanov}).
\begin{equation}
{t_\parallel } = \frac{{{L_\parallel }}}{\tilde{c}_+} + \frac{{{L_\parallel }}}{\tilde{c}_-} = \frac{{2Ln}}{c}\left[ {1 - \frac{{{V^2}}}{{{c^2}}}\frac{\Delta \varepsilon(1 - \Delta {\varepsilon})}{\varepsilon }} \right]\label{time parallel}
\end{equation}
in the form not needed any other corrections to calculation of times ${t_\parallel }$  and ${t_ \bot }$  by  geometrical optics. Specifically, the lag $t_\|$ needs not the correction for Lorentz contraction of the arm $L_\|$, and the lag in the arm ${L_ \bot }$ perpendicular to $\mathbf{V}$ with the account of ${V_ \bot } = 0$  does not demand the calculation of "Lorentz triangle" in order that to obtain in the outcome the formula (\ref{shift Demjanov}) which describes the experiment according to Fig.3. This is seen after applying Lorentz-invariant formula (\ref{time parallel}) for calculation the lag in the arm ${L_ \bot }$ perpendicular to $\mathbf{V}$ when ${V_ \bot } = 0$, and $\tilde{c}_\pm=c/n$:
\begin{equation}
{t_ \bot } = \frac{{{L_\perp}}}{\tilde{c}_+} + \frac{{{L_\perp}}}{\tilde{c}_-} = \frac{{2{L_ \bot }n}}{c}.\label{time perpendicular Michelson medium}
\end{equation}
Indeed, formula (\ref{time perpendicular Michelson medium}) coincides with the initial formula (\ref{time perpendicular Michelson}) of Michelson differing from (\ref{time perpendicular Michelson}), where $n=1$, only by the accounting the actual refractive index of the medium, $n>1$. Subtracting expression (\ref{time parallel}) from (\ref{time perpendicular Michelson medium}) gives for the difference of propagation times $\Delta t = {t_ \bot } - {t_\parallel }$ the final expression (for ${L_ \bot } = {L_\parallel } = L$):
\begin{equation}
\Delta t = {t_ \bot } - {t_\parallel } = \frac{{2L}}{{c\sqrt \varepsilon  }}\frac{{{V^2}}}{{{c^2}}}\Delta \varepsilon(1 - \Delta \varepsilon ).\label{time difference medium}
\end{equation}
Substituting (\ref{time difference medium}) into expression of the amplitude $A_m=c\Delta t/\lambda$  of the fringe shift as a function of $\Delta t$ gives (\ref{shift Demjanov}), therefrom follows (\ref{aether wind Demjanov}).

Thus, we have shown that using the invariant formula (\ref{expansion abridged}) for calculation of time $t_\|$ of light propagation in the arm of the interferometer parallel to $\mathbf{V}$  gives directly formula (\ref{shift Demjanov}) without corrections neither in the form of "Lorentz triangle" in the arm $L_\perp$ nor as Lorentz contraction of the arm $L_\|$. Such derivation of formula (\ref{shift Demjanov}) can be taken to meet the principle of relativistic Lorentz-invariance, and formula (\ref{shift Demjanov}) itself can be considered to be invariant in the above mentioned a new sense regarding all optical media   with any value of $c/n$  both moving and stationary.  The postulate of the relativistic theory of 20th century concerning the independence of the speed of light in moving or stationary inertial frames extends in this way to all optical media (in space regions where $n=\mathtt{const}$), and the constancy of the light speed is reasonable to say about only in the bounds of the medium where $n=\mathtt{const}$. Clearly, vacuum (aether with $n=1$) is rather a particular case, covered by special relativity, among a much more wide variety of dynamic manifestations of "relativity" of entities in aether.

\section{Conclusion}

Thus, if you have already performed experiments on Michelson interferometer with high optical density light's carriers (e.g., \cite{Shamir}), where for one reason or another there has not been recorded the shift of interference fringe, you should not rush to insist on the historically erroneous conclusions of science of early 20th century that Michelson-type experiments  are allegedly ''negative''. Instead you should look for methodical artifacts in the design of your apparatus and remove them, as I did by the above described method when faced with the instrumental origin of the absence of fringe shift in an interferometer with liquid and solid light's carriers. The above described means to eliminate artifacts in Michelson interferometer (Fig.\ref{fig1}), my experimental results (Fig.\ref{fig2} and Fig.\ref{fig3} ) and their interpretation \cite{Demjanov rus, Demjanov} testify to a great potential in increasing the interferometer's sensitivity to ''aether wind'' enabling us to achieve the ''positiveness'' of Michelson-type experiments, and thus to prove in principle the possibility to observe absolute motion of inertial bodies.

This experimental success cannot be considered separately from the mathematical methods and models of interpreting the experimental results. Mentioned in section 7 analysis shows how great is the influence of herediting the prior notions pertaining to pioneer explorers of this or another natural phenomenon on subsequent attempts to interpret new inevitably emerging experimental results. There needed 70 years in order that the Michelson supposition concerning the possibility of detecting aether wind in vacuum medium has been subjected to doubt \cite{Demjanov rus, Shamir, Trimmer, Cahill Kitto}, thoroughly rechecked \cite{Demjanov rus} and disproved. The disproof consisted in the experimental demonstrating \cite{Demjanov} that in order to detect aether and absolute motion it is necessary the translational motion of the particles of detecting medium to be engaged thus ensuring the refractive index $n>1$ at any instant of an optical medium moving in aether, the latter having the refractive index $n_\texttt{aether}=1$.

Now we know at least five means to interpret mathematically the experimental results of Fig.3 leading to formula (\ref{aether wind Demjanov}) determining the aether wind velocity directly by processing the observed fringe shifts. One of these means, above described in section 7.3, revealed the main cause of failure to interpret many of experimental measurements of aether wind velocity during such a long period of time since 1881. The cause consisted in that from the very beginning physical processes occurring in Michelson interferometer required nonclassical (relativistic) description, which could not be available in the end of 19th century. In the 5th and 6th versions of preprint I for the first time present such a description of the processes occurring in the device which from the very beginning are based on the applying tools of interpreting their relativistic Lorentz-invariance, which became known only in the end of 20th century. In this event I extend the notion of Lorentz-invariant descriptions of moving media to composite systems of simultaneously polarizing their parts $\varepsilon=1.+\Delta\varepsilon$, arising from stationary aether $\varepsilon_\texttt{aether}=1.$ and movable in it particles ($\Delta\varepsilon>0$).

The maximal speed of aether wind, obtained from the observation of the greatest shift of the interference shift at the latitude of the town Obninsk ($\sim 56^\circ$ N) in the horizontal plane of rotation of the device, was $V'\approx 480$ km/s. This is, as is known, agreed with the logic of well known astronomical observations of the motion of the Earth at some orbits of revolution (especially around the Sun and the center of Galaxy). Such a compatibility of physical experiments performed in different domains of physics attests to principal possibility of the observing (by means of the Michelson interferometer) absolute motions of inertially mobile objects relative to aether space.
%\enlargethispage{25pt}
%\pagebreak

\end{document}